\documentclass[preprint]{aastex701}

\usepackage{graphicx}
\usepackage{dcolumn}
\usepackage{bm}
\usepackage{amsmath}
\usepackage{soul}

\begin{document}

\title{Time-dependent cosmic-ray escape from wind bubbles: hard spectra formation}

\author[orcid=0000-0003-1332-9895,sname='Merten']{Lukas Merten}
\affiliation{University of Wuppertal, Gaußstrasse 20, 42119 Wuppertal, Germany}
\email[show]{merten@uni-wuppertal.de}  

\author[orcid=0000-0002-4777-4842,sname='Aerdker']{Sophie Aerdker} 
\altaffiliation{Ruhr University Bochum, RAPP Center}
\affiliation{Astroparticule et Cosmologie, Universite Paris Cit\'e, 10 Rue Alice Domon et L\'eonie Duquet, F-75013 Paris, France}
\email[show]{sophie.aerdker@rub.de}

\author[orcid=0000-0003-0543-0467, sname=Peretti]{Enrico Peretti}
\affiliation{INAF, Istituto Nazionale di Astrofisica, Osservatorio Astronomico di Arcetri, Largo E. Fermi 5, 50125 Florence, Italy}
\affiliation{Astroparticule et Cosmologie, Universite Paris Cit\'e, 10 Rue Alice Domon et L\'eonie Duquet, F-75013 Paris, France}
\email[show]{enrico.peretti.science@gmail.com}

\begin{abstract}

    Wind-driven bubbles are dynamic systems that can accelerate cosmic rays, depending on their physical properties, up to very high energies. We investigate how a time-dependent description of the particle transport may impact the escaping cosmic-ray flux. The wind bubble system is modeled as spherically symmetric. Cosmic rays are continuously injected at the position of the termination shock and propagate through advection and diffusion until the escape at the time-dependent position of the forward shock, which is treated as a free escape boundary. The one-dimensional spherical time-dependent transport equation is solved by transforming it into the corresponding set of stochastic differential equations, and integrated with a modified version of the open source cosmic-ray propagation framework CRPropa. We find that, during the wind driven phase, the downstream escaping spectra from wind bubbles can be harder than $\sim E^{-2}$, the conventional expectation from diffusive shock acceleration. Depending on the turbulence model the initial energy spectrum can be significantly suppressed at lowest energies, which could be an observable feature to distinguish between different turbulence realizations. This effect could lead to an efficient confinement of low energy particles, potentially leading to observable implication in terms of multi-messenger radiation and cosmic-ray accumulated grammage within the bubble.

\end{abstract}


\section{Introduction}
\label{sec:intro}

Wind-driven bubbles are observed across a wide range of spatial scales and powers throughout the Universe: from astronomical-unit scales in the environments of main-sequence stars to parsec scales around Wolf–Rayet stars; from tens of parsecs for bubbles inflated by star clusters up to hundreds of parsecs in the case of ultra-luminous X-ray sources. Even galactic winds, which extends for several kiloparsecs in galactic halos, may exhibit a bubble-like structure.

These systems are powered by relatively compact sources that drive winds with wide opening angles, and are characterized by a typical onion-like structure \citep{Weaver-etal-1977}. An inner wind termination shock separates the innermost fast, cool wind from the outer hot, shocked wind, while an outer forward shock propagates into the ambient medium. Behind the forward shock lies the shocked ambient gas, which is highly compressed in a thin shell and physically separated from the wind material by a contact discontinuity.
In the early stages of evolution, the forward shock and the wind termination shock are closely coupled, expanding together at approximately constant velocity. Once the swept-up mass becomes comparable to the total injected mass, the system enters a deceleration phase and the two shocks decouple. In this regime, the forward shock expands as 
$R_{\rm fs}\sim t^{3/5}$, while the termination shock follows $R_{\rm ts}\sim t^{2/5}$ \citep{Weaver-etal-1977,Koo92_a,Koo92_b}.

Wind bubbles are of particular interest because diffusive shock acceleration (DSA) at their shock fronts can energize particles up to very high energies. In a series of pioneering works applied to different sources~\citep{Morlino-etal-2021,Peretti_SBG,MukhopadhyayEtAl2023,Peretti_UFO,Peretti_ULX}, it has been suggested that efficient particle acceleration may occur at wind termination shocks under quasi-stationary conditions, thereby favoring the attainment of extremely high energies.
Different numerical simulations confirmed that wind termination shocks, rather than forward shocks, can be expected to be the most relevant site for particle acceleration via DSA \citep{Gupta2020,Meliani2024}.

In the context of Galactic cosmic rays (CRs), wind bubbles inflated by star clusters \citep{Bykov-Fleishman-1992,Bykov-etal-2020,Morlino-etal-2021,Vieu-etal-2022} and by super-Eddington accreting stellar-mass compact objects~\citep{Peretti_ULX,Pasquevich2026}, together with microquasar's jets~\citep{Abaroa-et-al-2024,Bykov2025,Zhang2025,Wang2025,Kaci2025,Peretti2026}, have emerged as promising candidates for accelerating particles up to the PeV range (so-called PeVatrons).
This possibility is especially timely given the challenges faced by supernova remnants~\citep{Cristofari2020}---long considered the primary candidates based on their energy budget~\citep{Blasi2013,Gabici2019}---in reaching such energies, as well as recent indications from LHAASO observations pointing at stellar clusters \citep{LHAASO2024-Cygnus} and accreting stellar-mass compact objects \citep{LHAASO2025-blackholes,LHAASO2026-Cygnus-X3}. 

Wind bubbles inflated by stellar clusters are particularly compelling as gamma-ray sources. They have been long observed by Fermi-LAT in the GeV domain~\citep{Fermi-LAT2011-Cygnus} and, recently, also at very-high-energies (VHE) by imaging air Cherenkov telescopes \citep{Aharonian2019}. 
Depending on the compactness of the cluster of stars inflating the bubble, there could be a strong wind termination shock~\citep{Gupta2020} or, in the case of loose clusters, no termination shock at all~\citep{Vieu-etal-2024}.
Supernovae exploding in star clusters environment might encounter ideal conditions for extremely efficient acceleration ~\citep{Bykov2018a,Bykov2018b,Vieu-etal-2022-II,Vieu2023}. Moreover, they may help explain isotopic anomalies (e.g., $^{22}$Ne) observed in cosmic-ray abundances~\citep{Casse1982,Tatischeff2021}.
Galactic stellar clusters have been classified according to their wind power~\citep{Celli2024} and assessed as potential sources of diffuse gamma rays and CRs in the Galaxy~\citep{Peron2024Nat,Peron2024Gamm-unass,Menchiari2025}, with numerous gamma-ray observations interpreted as either hadronic~\citep{Peron2025,Liu2022} or leptonic~\citep{Harer2023} emission; notably, one of the most powerful clusters, Westerlund 1, is observed to drive a strong outflow expanding beyond the galactic disk~\citep{Lemoine-Goumard2025}.

Although CRs can be efficiently accelerated at the inner wind termination shock---provided that the cluster is compact---the time evolution of wind bubbles and its impact on the escaping cosmic-ray flux injected into the interstellar medium remain poorly explored. This aspect may represent a crucial piece of information for assessing the contribution of this class of sources to the cosmic-ray spectrum observed at Earth, as well as to the gamma-ray emission from clusters and their surroundings.

Time-dependent treatments of acceleration and transport of particles at shocks in the Galactic halo showed that the spectrum at the shock can be altered depending on its geometry, and the observed spectrum at a given distance downstream can be significantly hardened depending on the considered diffusion coefficient and observing time \citep{Drury, shock-landscape}. 

In this work, we aim to investigate the time-dependent transport of CRs in a wind bubble. In particular, we focus on the scientific case and parametric configuration typical of star clusters. Nevertheless, our findings can be easily extended to wind bubbles of different powers and scales. 
Particles accelerated at the termination shock are advected downstream until they eventually reach the forward shock, typically considered as the escape boundary into the Interstellar medium. The distance between the two shocks increases over time, making it harder for the particles to escape the system. In the presence of energy-dependent diffusion, high-energy particles are more likely to escape, while low-energy particles can be trapped in the bubble for a sizable amount of time. Such a confinement might result in relevant multi-messenger emission. Energy losses of low energy particles might lead to a non-negligible confinement preventing these particles to eventually escape the system.

To study particle transport in the time evolving bubble we integrate stochastic differential equations (SDEs), that are equivalent to a Fokker-Planck equation. 
SDEs have been successfully used to model e.g.~time-dependent CR transport in the Galaxy \citep{Merten-etal-2018}, DSA at stationary shocks \citep{Aerdker-etal-2024}, and in the case of time-dependent shocks in the Galactic halo \citep{shock-landscape}. They provide a fast and flexible method complementary to grid-based methods \citep{StrongMoskalenko98, Evoli-etal-2008, Kissmann-2014} that can easily be extended to account for the time evolving systems at hand or different stochastic drivers \citep{Effenberger-etal-2024, Aerdker-etal-2025}. SDE models are also well-established to describe particle transport in the heliosphere \citep[see, e.g.][for an overview]{Strauss-Effenberger-2017}. 
The SDEs are integrated with a modified version of the open-source framework CRPropa3.2 \citep{GeneralSDE, CRPropa3.2}. This extended version allows for diffusive transport in the presence of time-dependent background fields, spatially changing eigenvalues of the diffusion tensor and comes with the infrastructure to keep track of individual phase-space elements which is beneficial for re-weighting.

We find that wind bubbles can release hard spectra of CRs into the interstellar medium  in certain physical configurations. We present results for three different diffusion scenarios: Kolmogorov, Kraichnan, and Bohm.

The manuscript is organized as follows. A detailed description of the model and numerical implementation is given in section~\ref{sec:methods}. Section~\ref{sec:results} describes the results, which are discussed and interpreted in section~\ref{sec:discussion}. This work concludes with a summary and outlook in section~\ref{sec:summary}.

\section{Methodology}
\label{sec:methods}

We present a time-dependent model of the cosmic-ray transport in a wind bubble environment. Both, the source of high-energy CRs, the termination shock, and the escape boundary into the Interstellar medium, the forward shock, move with different, time-dependent speed. 
This makes it necessary to go beyond stationary descriptions, which might not capture important physics implications as shown later.

We employ a spherical symmetric set up of the environment as detailed in section \ref{ssec:setup}, based on which we estimate the advection and diffusion time scales in \ref{ssec:time-scales}, and use methods of stochastic differential equations to solve the time-dependent transport equation of CRs (see section \ref{ssec:model}).

\subsection{Time dependence of termination and forward shock and wind profile}
\label{ssec:setup}

Throughout this work, we assume a one-dimensional spherical system. Three-dimensional MHD simulations of compact young massive stellar clusters (YMSCs) show that the termination shock and flow might deviate from such a uniform assumption, depending on the distribution of stars in the cluster \citep{Haerer-etal-2025}. However, the general evolution of the bubble is consistent with a 1D treatment, and as we focus on the time-dependency of the system, a combination with a 3D treatment in space remains subject to future work.


The radii of the termination and forward shocks expanding in a medium of density $n_0$ over time can be estimated based on the mass loss rate $\dot{M}$ and the terminal velocity $v_\infty$ of the collective wind of the bubble \citep{Weaver-etal-1977}. The deceleration radius can be defined as the radius at which the swept-up mass balances the mass of the wind material, $(4 \pi/3) R_{\rm dec}^3 m_\mathrm{p} n_0 = \Dot{M} t_{\rm dec} $, where $t_{\rm dec}= R_{\rm dec} V_{\infty}^{-1}$ is the deceleration time. With the analytical expression for the deceleration time, the forward shock radius can then be expressed as $R_{\rm fs} = R_{\rm dec} (t/t_{\rm dec})^{3/5}$ while the termination shock $R_{\rm ts} = R_{\rm dec} (t/t_{\rm dec})^{2/5}$  \citep[see e.g.][]{Weaver-etal-1977}.
Adopting the expression of deceleration time, we obtain the following expression for the forward shock
\begin{align}
    &R_{\rm fs} = \left( \frac{3}{2\pi} \frac{\Dot{M}V_{\infty}^2}{2} \frac{1}{m_p n_0} \right)^{1/5} t^{3/5} \nonumber \\ 
    & \approx 80 \, {\rm pc} \, \left(\frac{\Dot{E}}{10^{38} \rm erg \, s^{-1}}\right)^{1/5} \, \left(\frac{n_0}{1 \,  \rm cm^{-3}}\right)^{-1/5} \, \left(\frac{t}{1 \, \rm Myr} \right)^{3/5}, \label{eq:forwardshock}
\end{align}
where $\Dot{E} = \Dot{M}V_{\infty}^2/2$. 
The termination shock is found similarly as
\begin{align}
    R_{\rm ts} &= \left( \frac{3}{4\pi} \frac{\Dot{M}}{n_0m_p} \right)^{3/10} V_{\infty}^{1/10} t^{2/5} \nonumber \\ 
    & \approx 15.8 \, {\rm pc} \, \left(\frac{\Dot{M}}{10^{-4} \rm M_{\odot} \, yr^{-1}}\right)^{3/10} \, \left(\frac{V_{\infty}}{10^{3} \,\rm km \,s^{-1}}\right)^{1/10} \left(\frac{n_0}{1 \,  \rm cm^{-3}}\right)^{-3/10} \, \left(\frac{t}{1 \, \rm Myr} \right)^{2/5} \quad . \label{eq:terminationshock}
\end{align}

Both shocks move with a different speed, $\dot{R}_\mathrm{ts} \propto t^{-3/5}$, $\dot{R}_\mathrm{fs} \propto t^{-2/5}$. The distance between the shocks increases over time, which makes a time dependent treatment of the escaping CRs necessary to accurately capture all spectral features.

We assume an effective time-evolving background flow constructed as a sequence of quasi-stationary snapshots. At each time step, the shocked wind region between the termination shock and the forward shock is described by a radial velocity profile, $u(r,t)\propto r^{-2}$, normalized to the post-shock velocity at the termination shock, $u(R_{\rm TS})\simeq V_\infty/4$. Such a profile approximately preserves spherical mass conservation and provides an effective description of particle advection within the bubble interior.
While the adopted flow does not correspond to a fully self-consistent solution of the time-dependent Euler equations, it represents an effective advection field experienced by CRs, capturing the dominant transport scales relevant for their propagation within the shocked wind region.

\subsection{Time scale arguments}
\label{ssec:timescales}

A qualitative intuition of the fate of cosmic-ray accelerated in wind bubbles can be inferred by comparing the typical timescales, especially those connected to their residency in the downstream region, where they are expected to be confined most of the time.

While diffusion might be relevant for high-energy particles and their escape, the advection is likely the mechanism regulating the transport and escape of particles at low energy. 
In particular, the advection timescale mediated over the downstream volume and for a wind profile $u(r) \propto r^{-2}$ reads
\begin{align}
    \tau_{\rm adv} = \frac{(R_{\rm fs}^3 - R_{\rm ts}^3)}{3 u_2 R_{\rm ts}^2} \lesssim \frac{R_{\rm fs}^3}{3 u_2 R_{\rm ts}^2} \sim t. 
\end{align}
Substituting the self-similar scaling of the shock radii during the deceleration phase it is straightforward to see that $\tau_{\rm adv} \sim t$.
Note, that the timescale estimated above assumes that the position of the forward shock is "frozen" at the time the particle was released.
The qualitative matching between the advection time---which is expected to regulate the escape of low-energy particles from the bubble---and the dynamical time of the system ($t_{\rm dyn} =R_{\rm fs}/\dot{R}_{\rm fs}$) may imply that low-energy particles might be trapped inside the bubble for a sizable fraction of the bubble lifetime. Later we will show that particles actually would be trapped indefinitely in the downstream region when the assumption of a "frozen" escape boundary is dropped (see the appendix).

\subsection{Time-dependent SDE model}
\label{ssec:model}

 In the test particle limit, the time evolution of the distribution function $f$ can be modeled by a spherical one-dimensional transport equation of the form \cite{Schlickeiser-89-I} 

\begin{align}
    \label{eq:transport}
    \frac{\partial f}{\partial t} &= \nabla \cdot \left(\kappa \cdot \nabla f - \mathbf{u} f\right) + \frac{p}{3}\nabla \cdot \mathbf{u} \frac{\partial f}{\partial p} + S
\end{align}
Here $\mathbf{u} = u(r,t)\mathbf{e}_r$ describes the time dependent advection, $\kappa(r, p, t)$ is the only non-zero diffusion coefficient depending on particle momentum and magnetic field properties $B(r, t)$, as well as sources and sinks given by $S(r, p, t)$. 
Diffusion in momentum (second-order Fermi acceleration) and any continuous losses are not considered in this paper but can be considered in future work.

The time-dependency of $u$ makes it difficult to solve equ.~\ref{eq:transport} analytically or on a grid. Previous work is often based on the derivation of transport time scales instead of solving the time dependent transport equation directly \citep[see, e.g.,][]{Haerer-etal-2023}. However, since equ.~\ref{eq:transport} is a Fokker-Planck equation, it can be re-written into a set of SDEs, following It$\hat{\mathrm{o}}$'s lemma. To avoid weighting terms, corresponding to terms linear in $f$, we introduce the quantity $\tilde{f}=fr^2$ and derive the corresponding SDE for $\tilde{f}$ instead of $f$ (see the appendix).
\begin{align}
\label{eq:SDEs}
    \mathrm{d}r &= \left( u(r, t) + \frac{\partial \kappa}{\partial r} + \frac{2\kappa}{r} \right)\mathrm{d}t + \sqrt{2\kappa} \mathrm{d}W_t
\end{align}
where $\mathrm{d}W_t$ is a Wiener process, modeling the random walk of CRs in turbulent magnetic fields. We omitted the second, ordinary differential equation, which describes momentum gains and losses, because our choice of the velocity field is divergence free (see sec.~\ref{sssec:advection}). The SDE corresponding to the distribution function $f$, including the weights, can be found e.g. in \citep{Kopp-etal-2012}.

Equation~\ref{eq:SDEs} models the time evolution of phase-space trajectories --- or samples of the distribution function $\tilde{f}$ --- not real particle trajectories. The distribution function $f$ is obtained by solving for many of such phase-space trajectories and creating a histogram of the pseudo-particle positions, including necessary weight terms $w\propto r^{-2}$. Sources and sinks are modeled by applying weights to individual phase-space trajectories. 

Here, we focus on time-dependent transport and escape of accelerated particles and do not explicitly model time-dependent acceleration at the termination shock. Instead, particles are injected at the position of the termination shock with a pre-defined energy-spectrum and high-energy cut-off depending on the time and size of the wind bubble (see sec.~\ref{sssec:injection}). 

Furthermore, particles are not adiabatically cooled in the downstream region since we assume $u(r > R_\mathrm{TS} \propto r^{-2})$ and spend only short times in the upstream region $u(r < R_\mathrm{TS})$ (see sec.~\ref{sssec:advection}). Thus, we neglect energy changes and focus on the time-dependent particle transport modeled by equ.~\ref{eq:SDEs} (see the appendix for a justification). The SDE is solved with the Euler-Maruyama scheme as described in \citep{GeneralSDE}.

For modeling the time-dependent escape of wind bubbles, the advection $u$ and diffusion $\kappa$ must be specified. The next sections present our time-dependent models, followed by a discussion of the injected spectrum at the shock, including a time-dependent maximal energy.

\subsubsection{Advection}
\label{sssec:advection}

The advection field forcing the particles to drift outwards is defined by a constant flow velocity before the termination shock $v_0$ and a mass conserving flow after the termination shock $u\propto r^{-2}$
\begin{align}
    {u}(r) = v_\infty \cdot \left( 1 + \left(\left(\frac{R_\mathrm{ts}}{2r}\right)^2-1\right) \frac{1}{1+\exp(-\frac{r-R_\mathrm{shock}}{l_\mathrm{ts}})}\right) \quad . \label{eq:advectionField}
\end{align}
The shock is smoothed out exponentially with a widths of $l_\mathrm{shock}$ that is small compared to the diffusive length scale of the particles. While the shock width has negligible impact on the transport of high-energy CRs, it becomes a relevant parameter when studying time-dependent acceleration as well \citep{KruellsAchterberg94, Aerdker-etal-2024}.

The position of the termination shock is weakly time dependent $R_\mathrm{ts}(t) = R_{\rm ts,0} \left(t / t_0\right)^{2/5}$, where $R_{\rm ts,0}$ is the termination shock position at $t = t_0$, the time at which our simulation starts.

The forward shock, located at $R_\mathrm{fs}(t) = R_{\rm fs,0}(t/t_0)^{3/5}$, is treated as free escape boundary, which means that the advection field for $r>R_{\rm fs,0}$ is not relevant for this work. 

This advection profile (equ. \ref{eq:advectionField}) extends earlier descriptions of the particle transport by modeling the position of the source and escape boundary in time consistently --- a particle ejected at time $t_1$ will need to travel further out than $R_\mathrm{fs}(t_1)$ since the free escape boundary will have moved. However, the profile between the shocks is only an approximation and not a self-consistent solution of the underlying hydrodynamic equations. This would require a coupling of the CRPropa framework with an HD solver, which is beyond the scope of this work.

\subsubsection{Magnetic field and diffusion}

The magnetic field plays a crucial role in the applied transport model as the diffusion coefficient scales with the local magnetic field strength and turbulence. 
The magnetic field strength is estimated assuming that in the upstream region of the termination shock it is characterized by a pressure equal to a fraction $\epsilon_B$ of the kinetic energy density, which is half of the ram pressure
\begin{equation}
    \frac{B_{\rm ts}^2}{8 \pi} = \epsilon_B \frac{1}{2} \rho u^2 = \epsilon_B \frac{1}{2} \frac{\Dot{M}}{4 \pi R_{\rm ts}^2 V_{\infty}} V_{\infty}^2 \quad . \label{eq:Bfield}
\end{equation}
For simplicity and because particles do not spend much time in the region it was assumed that $B$ is spatially constant in the upstream region. In the downstream region the magnetic field is amplified by a factor $\sqrt{11}$ assuming isotropic turbulence, and is radially decreasing behaving with a given power law index $\alpha$. In particular, we explore the scenarios $\alpha=0,1,2$.
In summary this leads to
\begin{align}
    B_u = \sqrt{\epsilon_B \Dot{M} V_{\infty}} \, R_{\rm ts}^{-1} \quad \text{and} \quad B_\mathrm{d} = \sqrt{11} B_\mathrm{u} \left(\frac{r}{R_{\rm ts}}\right)^{\alpha} \quad . \label{eq:bfield}
\end{align}
Following quasi-linear theory we specify the diffusion coefficient
\begin{align}
    D = \frac{1}{3} \frac{v(p) r_\mathrm{L}}{\mathcal{F}(k)},
\end{align}
where $v(p)$ is the momentum-dependent particle velocity. Here, we assume the relativistic limit $v(p)=\mathrm{c}$ as we limit the simulations to energies above $E_\mathrm{min}=5\,\mathrm{GeV}$. Furthermore, $r_L$ is the Larmor radius and $\mathcal{F}(k)$ is the normalized energy density per unit of logarithmic wave number; practically the normalized turbulence power spectrum, which can be approximated as $(l_c r_L^{-1})^{1-\delta} $ so that the diffusion coefficient can be re-expressed as follows 
\begin{align}
    D(r,E) \approx \frac{c}{3} r_\mathrm{L}^{2-\delta}(r,E) l_\mathrm{c}^{\delta-1} .
\end{align}
Note, that the latter equation holds for $r_\mathrm{L} \lesssim l_\mathrm{c}$ and has to be adjusted to the quasi-ballistic \citep{ReichherzerEA2020, ReichherzerEA2022} regime for $r_\mathrm{L} > l_\mathrm{c}$
\begin{align}
    D(r,E) = \frac{cl_\mathrm{c}}{3} \left[\frac{r_\mathrm{L}(r,E)}{l_\mathrm{c}}\right]^2 \quad .
\end{align}
However, for the parameters studied in this work the Larmor radius $r_\mathrm{L}$ is usually smaller than the correlation length $l_\mathrm{c}$.

\subsubsection{Injection at the shock}
\label{sssec:injection}

Particles are injected at the termination shock assuming that they are accelerated to a power law spectrum with a spectral index of $\gamma = -2$ (in energy) and a cut-off due to the maximal energy that can be reached at a given time. 

The phase space is sampled for energies $E \in [5\,\mathrm{GeV}, 10\,\mathrm{PeV}]$ and a power-law of $E^{-1}$, to ensure same statistics in energy bins $\mathrm{d}N/\mathrm{d}E$. After simulation, weights are applied to 1) weigh the spectrum to $\mathrm{d}N/\mathrm{d}E \propto E^{-2}$, and 2) apply additional weights to adjust for the cut-off. 
The shape of the cut off is given by
\begin{align}
    w = \left[1 + \alpha_1 \left(\frac{E}{E_\mathrm{max}}\right)^{\alpha_2} \right] \cdot \exp{\left(-\alpha_3 \left(\frac{E}{E_\mathrm{max}}\right)^{\alpha_4} \right)} \quad , \label{eq:cutoff}
\end{align}
which is taken from \cite{Menchiari-etal-2024}. The total weight is therefore given by $w_\mathrm{tot} = w \cdot E_0^{-1}$. The more complex form compared to a pure exponential cut off comes from taking the spherical geometry and acceleration process into account. The parameters $\alpha_i$ depend on the realized diffusion (Bohm, Kraichnan, and Kolmogorov) and are given in appendix~\ref{Appendix: Emax}.
The maximal energy is calculated from the acceleration time and a Hillas limit. Details on how the maximal energy is calculated are also stated in appendix~\ref{Appendix: Emax}. 
The source moves outwards with the termination shock and its luminosity remains constant over time.

The simulation starts at $t_0 = 0.01\,\mathrm{Myr}$ when the deceleration phase sets in. At this time, the shocks are already separated by $\sim1.5\,\mathrm{pc}$ and injecting particles only at the termination shock from $t_0$ on neglects those that could have been accelerated at earlier times and are now diffusing downstream between the shocks. Thus, a convergence phase is added, where particles are injected between the two shocks ($r_\mathrm{ts}\leq r_\mathrm{source} \leq r_\mathrm{fs}$ for $t < t_0$) before the shocks start to move. At $t = t_0$, thus, also low energy particles can escape the system as shown in the appendix. However, once the forward shock moves, the diffusive length scale of low energy particles is too small, preventing them from escape. The convergence phase has a negligible influence on the escaping spectra at later times. 

\subsubsection{Parameters}

Simulations were performed for different turbulence models: Kolmogorov ($\delta = 5/3$), Kraichnan ($\delta = 3/2$), Bohm ($\delta = 1$), and fraction $\epsilon_B=(0.01, 0.1)$ of the ram pressure that goes into the magnetic field. Furthermore, different radial dependencies of the downstream magnetic field were tested, with $\alpha=(0, -1, -2)$. 

The other parameters were fixed throughout the simulations. The mass loss rate $\dot{M}=10^{-4}M_\odot\,\mathrm{yr}^{-1}$, the terminal velocity of the wind in the upstream region $v_\infty = 3\times 10^3\,\mathrm{km}\,\mathrm{s}^{-1}$, the correlation length of the magnetic field turbulence $l_c=2\,\mathrm{pc}$, and the density of the ambient medium $n_0=10\,\mathrm{cm}^{-3}$.

This leads to the following initial values of the two shocks $R_\mathrm{ts,0}=1.4\,\mathrm{pc}$ and $R_\mathrm{fs,0}=3.9\,\mathrm{pc}$; based on the evaluation of equ.~\ref{eq:terminationshock} and equ.~\ref{eq:forwardshock}, respectively, at the start of the simulation $t_0=0.01\,\mathrm{Myr}$.

\subsection{Technical implementation}

This section gives a brief overview on the technical implementation of the model in the modified CRPropa version described in \citep{GeneralSDE}. In this work CRPropa's candidates sample the scaled phase space density $\tilde{f}$ by solving the SDE given in equ.~\ref{eq:SDEs}; the deterministic coefficient is given by $A=u(r, t) + \partial \kappa / \partial r + 2\kappa/r$ and the stochastic one by $B=\sqrt{2\kappa}$.

Continuous injection at the termination shock is modeled by drawing a random time $t_\mathrm{inj}$ uniformly distributed between $t_0$ and 5~Myr. The pseudo-particle's source position is than calculated from the termination shock position $r_\mathrm{src}=R_\mathrm{TS}(t_\mathrm{inj})$. 

The free escape boundary is implemented through a testing if the pseudo-particle's current position at a given time $t$ is larger than the forward shock radius; if $r(t)>R_\mathrm{FS}(t)$ the candidate is deactivated and all its properties are stored. This leads to overshooting of the escape boundary. However, we have found that this is for the chosen time step $h=0.1\,\mathrm{kyr}$ not a problem. Although very few pseudo-particles might overshoot the boundary significantly the vast majority (90\% of the population) is usually observed with less than 3\% deviation from the shock position.

The simulations are based on the injection of $1.5\times 10^7$ pseudo particles per simulation set up. The Monte Carlo error, shown in the figures of the energy spectrum, suggests that this number is large enough.

\section{Results}
\label{sec:results}

We focused on one of the potential observables in this work: The energy spectrum at the free escape boundary; both at early ($t_\mathrm{obs}=(0.01, 0.015, 0.03, 0.1)\,\mathrm{Myr}$) and late times ($t_\mathrm{obs}=(0.1, 0.5, 1, 2.5, 5)\,\mathrm{Myr}$). We limit our simulation to 5 Myr, as the wind power is expected to decline thereafter due to the onset of supernova explosions~\citep{Vieu-etal-2022}.

\subsection{Energy spectra}

Energy spectra are derived following \citep{GeneralSDE} and references therein.

At early times the maximum energy of escaping CRs is increasing over time due to effects discussed in section~\ref{sssec:injection}. 
In addition, low energy CRs are depleted over time. This is due to the fact that particle transport, dominated at low energies by advection, is too slow to keep up with the escape boundary moving away with the speed of the forward shock. At higher energies particles can still reach the boundary due to their larger diffusion coefficient (see fig.~\ref{fig:escape_spectrum_alpha=0_early} for an overview).

Generally, the depletion effect at low energies is most pronounced for Bohm diffusion (smaller diffusion coefficient) and almost absent for Kolmogorov diffusion, while the Kraichnan scenarios lies somewhere in between (see also the figures in the appendix).

Figure~\ref{fig:lateBohm} shows exemplarily the time evolution of the escaping spectrum in a scenario of Bohm diffusion ($\delta=1$), constant downstream magnetic field ($\alpha=0$), and strong magnetization ($\epsilon_\mathrm{B}=0.1$). It is clearly visible that low energy particles with ($E<1\,\mathrm{TeV}$) are completely depleted and remain trapped in the downstream region between the two shocks. Above $\sim 10$ TeV the escaping spectrum is slightly harder than the injected one and the high energy cut off is stationary. 

\begin{figure}[htbp]
    \centering
    \includegraphics[width=.75\linewidth]{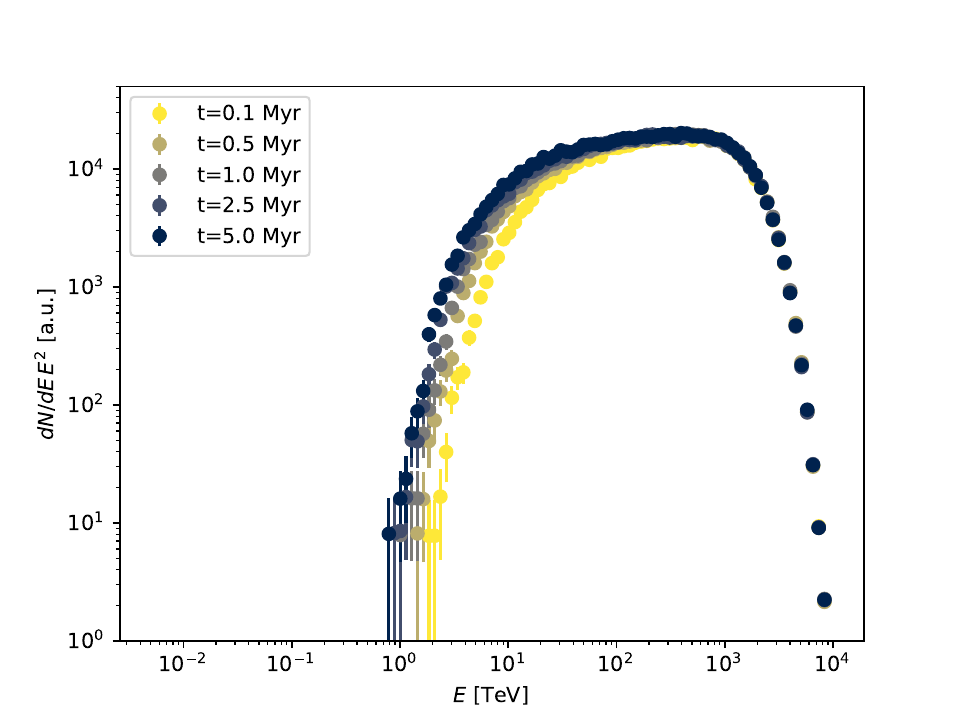}
    \caption{Escaping CR flux weighted with the $E^2$ for a constant downstream magnetic field strength ($\alpha=0$), Bohm diffusion coefficient ($\delta=1$) and large magnetization ($\epsilon_B=0.1$). The time evolution is color coded with bright yellow corresponding to earlier and dark blue colors corresponding to later times.}
    \label{fig:lateBohm}
\end{figure}

\begin{figure}[htbp]
    \centering
    \includegraphics[width=.75\linewidth]{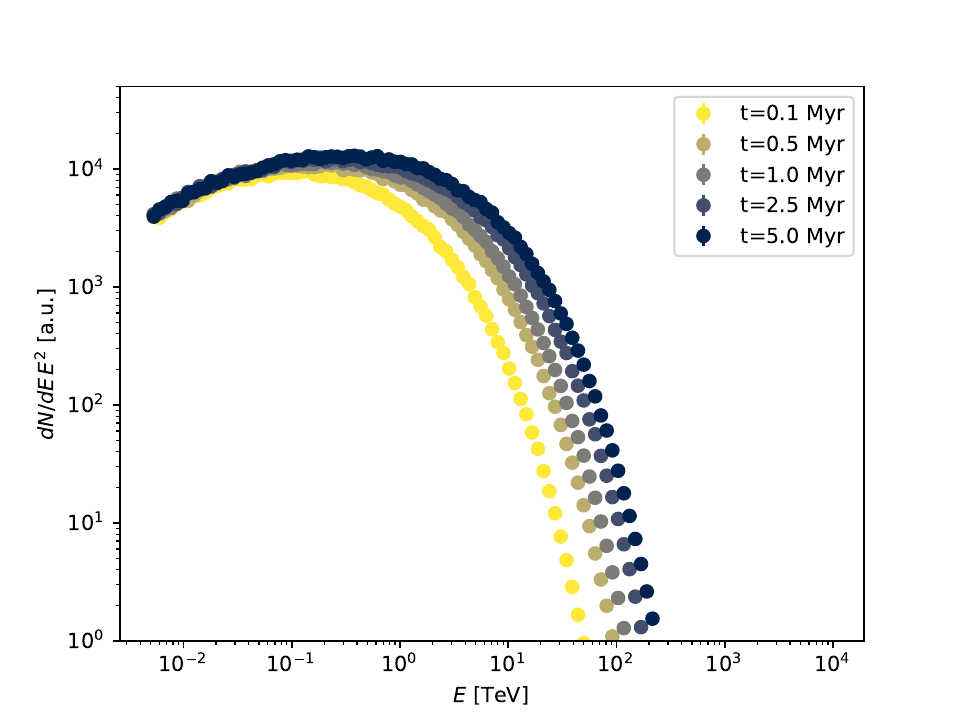}
    \caption{Same as fig.~\ref{fig:lateBohm} but for Kraichnan diffusion ($\delta=3/2$). While the low energy depletion is stationary the maximal energy is still increasing over time.}
    \label{fig:lateKraichnan}
\end{figure}

Figure~\ref{fig:lateKraichnan} shows the escaping spectra for Kraichnan diffusion ($\delta=3/2$). As the diffusion coefficient is generally larger than for Bohm diffusion the maximal energies are smaller and the low energy depletion is less pronounced. 

\begin{figure}[htbp]
    \centering
    \includegraphics[width=.75\linewidth]{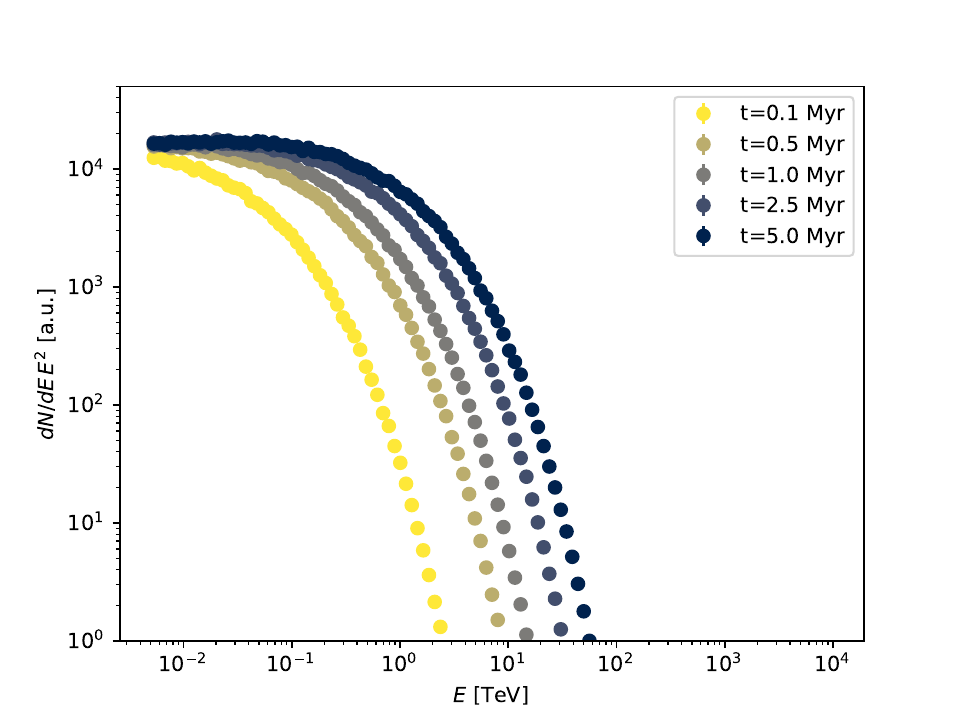}
    \caption{Same as fig.~\ref{fig:lateBohm} but for Kolmogorov diffusion ($\delta=5/3$). The maximal energy is strongly time dependent, but the low energy depletion has almost vanished.}
    \label{fig:lateKolmogorov}
\end{figure}

Figure~\ref{fig:lateKolmogorov} shows the escaping spectra for Kolmogorov diffusion ($\delta=5/3$). The even larger diffusion coefficient, compared to Kraichnan diffusion, prevents the depletion of low energy particles almost completely. The maximum energy is now strongly time dependent.

If and when the depletion at low energies or the maximal energy become stationary strongly depends on the model parameters (see the appendix for a complete overview). In section \ref{sec:discussion} we further estimate the low energy limit for escape based on the fastest particles that diffuse out of the system. 

In general, the slope of the escaping spectrum is determined by the energy-dependence of the diffusion coefficient, while the low-energy cut-off depends on its normalization. The influence of individual parameters is discussed in the following sections. 

\subsubsection{Influence of the magnetic field dependence}

While both the up- and downstream magnetic field strength at the termination shock are determined by the dynamics of the wind model itself (compare equ.~\ref{eq:bfield}), different radial dependencies of the downstream magnetic fields strength between the two shocks are plausible. We modeled the CR transport for three different values of the spectral index $\alpha=[-2, -1, 0]$. In general, a decreasing magnetic field ($\alpha < 0$) leads to a higher diffusion coefficient, thereby a more efficient escape from the system. This allows more CRs to reach the moving forward shock and contribute to the observed CR flux.

Figure~\ref{fig:comparison} shows a comparison of the influence of the radial dependence for the case of Bohm diffusion ($\delta=1$) and low magnetization ($\epsilon_\mathrm{B}=0.01$). It can be clearly seen that the low energy depletion of CRs is stronger for a constant magnetic field. The low energy cut off of the spectrum is decreased by roughly one order of magnitude when the radial spectral index $\alpha$ is reduced by one. Since the flux at the high energy cut off---which is independent of the downstream magnetic field---is almost constant, this results in harder spectra for weaker magnetic field dependence. 

\subsubsection{Influence of the diffusion model}

\begin{figure}[htbp]
    \begin{minipage}{.49\linewidth}
        \includegraphics[clip=True, width=\linewidth]{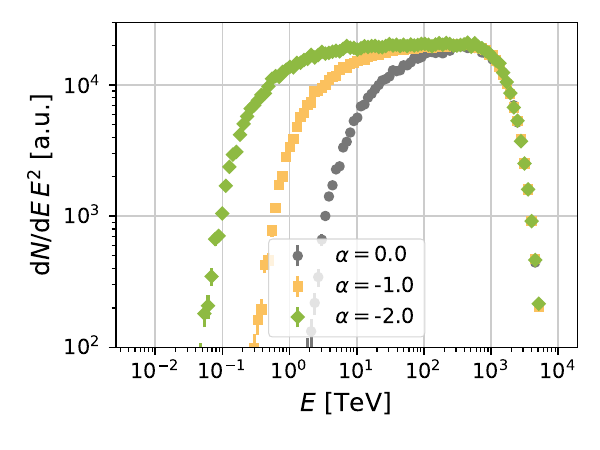}
    \end{minipage}
    \begin{minipage}{.49\linewidth}
        \includegraphics[width=\linewidth]{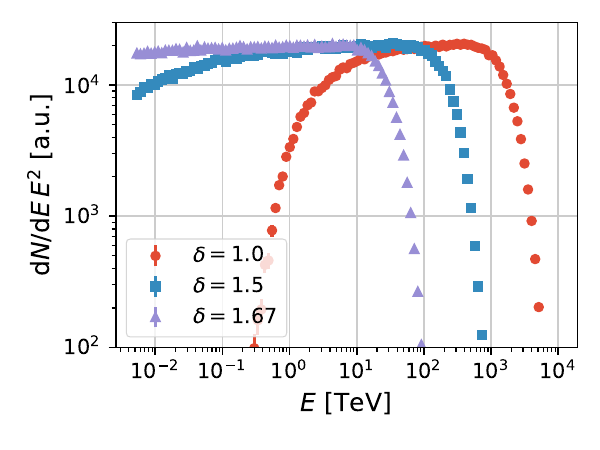}
    \end{minipage}
    \caption{Influence of different model parameters on escaping spectra at $t=1\,\mathrm{Myr}$ for large magnetization ($\epsilon_\mathrm{B} = 0.1$). (Left) Different radial dependencies (fixed to Bohm diffusion) (Right) Different diffusion models (fixed to constant magnetic field).}
    \label{fig:comparison}
\end{figure}

The actual turbulence characterizing wind bubble systems around YMSCs is not known. Therefore, the three most prominent homogeneous turbulence assumptions, namely Bohm ($\delta=1$), Kraichnan ($\delta=3/2$), and Kolmogorov diffusion ($\delta=5/3$), have been tested in this work. The chosen turbulence model influences the escaping spectral shape in two ways: 1) Following the arguments presented in \cite{Morlino-etal-2021,Menchiari-etal-2024}, the form of the high energy cut-off as well as the maximal energy are determined by the turbulence model. 2) In addition, also the low energy depletion strongly depends on the energy dependence of the turbulent cascade $\delta$ as can be seen in fig.~\ref{fig:comparison}. For a larger energy dependence, i.e. Bohm diffusion, the depletion is much more pronounced as for Kraichnan and Kolmogorov diffusion. For the latter one, the escaping spectrum is only slightly harder compared to the one injected at the termination shock. Kraichnan diffusion leads to an average hardening of about $\Delta\gamma=\gamma_\mathrm{esc}-\gamma_\mathrm{injected}=0.1-0.2$, depending on the magnetic field radial dependence $\alpha$ and the magnetization $\epsilon_B$. 
This means that an energy spectrum typical of DSA at the termination shock, $\mathrm{d}n/\mathrm{d}E\propto E^{-2}$, would roughly be observed when escaping as  $\mathrm{d}n/\mathrm{d}E\propto E^{-1.8}$ before the high energy cut off. 

Note, that in general the energy spectrum changes smoothly with energy, making it hard to fit a power law spectral index. 

\section{Discussion}
\label{sec:discussion}

We conclude that there are physically motivated configurations for which the energy spectra of escaping particles are significantly altered from those injected at the termination shock, due to a time-dependent treatment of the transport and the escape boundary. We emphasize again that this is only due to the time resolved treatment of the transport and not related to any loss processes, which are not taken into account in this work. 
If superbubbles contribute to the observed CR flux they could contribute with hard spectra. In optimistic parametric configurations, the hard spectra might come with almost no escaping CRs at low energies. 
This conclusion might be relaxed in a mixed diffusion model: where acceleration is driven by Bohm diffusion around the termination shock and the transport is dominated by Kraichnan or Kolmogorov diffusion.

On the other hand, pure Kolmogorov diffusion at the termination shock and throughout the superbubble, does not influence the escaping spectra significantly. However, CR hardly exceed a maximal energy of $\sim 20\,\mathrm{TeV}$ in this scenario. 

In the following, we discuss those results further, estimating the arrival times of the escaping high-energy particles in section \ref{ssec:arrivaltimes}. The fate of low energy particles is briefly discussed in the outlook (see section~\ref{sec:summary}.

\subsection{Transport to Earth}
\label{ssec:arrivaltimes}

With a time-dependent treatment of the source itself, it is interesting to estimate arrival times for different sources in our Galaxy. Further, the spectrum is changed due to energy-dependent diffusion in the Galactic magnetic field. As the transport in the Galaxy is a complex topical challenge \citep[see e.g.][]{Merten-etal-2017, Meinert_ICRC_2025} we only take a closer look at two limiting cases: 1) Transport along a flux tube, which corresponds to a vanishing perpendicular diffusion coefficient $\kappa_\perp=0$ and 2) 3D isotropic transport. The two Green's functions read
\begin{align}
   f(r,p,t)_\mathrm{FT} &= \frac{1}{\sqrt{4 \pi \kappa(p) t}} \mathrm{exp}\left[ - \frac{r^2}{4\kappa(p_0)t}\right] \quad \text{and} \\
   f(r,p,t)_\mathrm{3D} &= \frac{1}{[{4 \pi \kappa(p) t}]^{3/2}} \mathrm{exp}\left[ - \frac{r^2}{4\kappa(p_0)t}\right] \quad ,
\end{align}
where $r$ is the distance between the super bubble and Earth and $p_0$ the particles momentum. Any energy losses are neglected. For the diffusion coefficient, the Galactic constant value $\kappa = 5\times10^{24}\mathrm{m^2/s}\,(E/\mathrm{GeV})^{1/3}$ is used. 

\begin{figure}[htbp]
    \centering
    \begin{minipage}{.49\linewidth}
        \includegraphics[width=\linewidth]{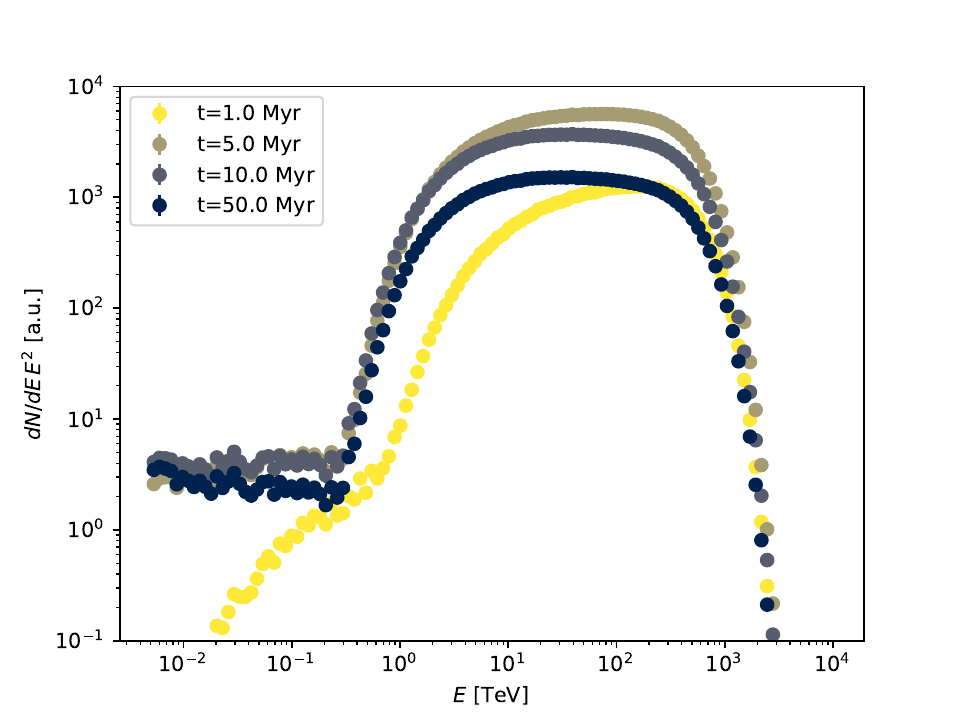}
    \end{minipage}
    \begin{minipage}{.49\linewidth}
        \includegraphics[width=\linewidth]{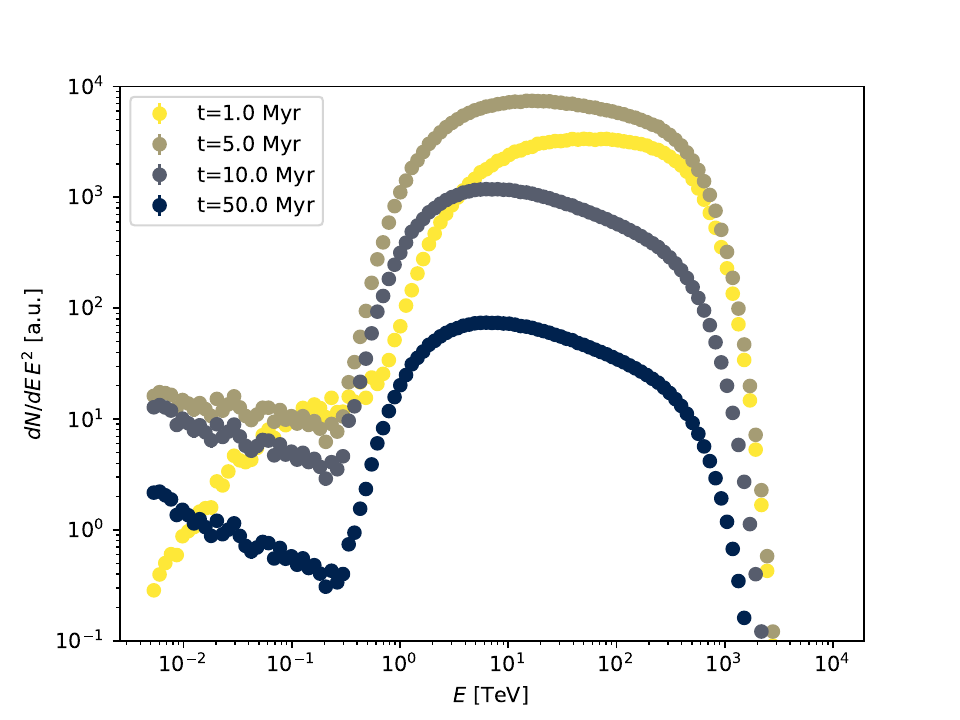}
    \end{minipage}
    \caption{Expected spectrum of CRs observed at Earth for a wind bubble that is located at a distance of $r=3\,\mathrm{kpc}$ in the case of pure parallel transport within a flux tube (left) and pure isotropic transport in 3D (right)}.
    \label{fig:earth}
\end{figure}

Exemplarily, we show the spectrum as it would be observed for a wind bubble at a distance of $r=3\,\mathrm{kpc}$ for different times after the wind was launched. The results are shown for the case of constant magnetic field ($\alpha=0$), Bohm diffusion ($\delta=1$), and low magnetization ($\epsilon_\mathrm{B}=0.01$) in fig.~\ref{fig:earth}. It can be seen that the low energy suppression is still visible at all times. The hardening of the spectrum is, however, partially smoothed out for later observation times; the softening of the spectrum is stronger in the 3D isotropic transport model. 

For other diffusion scenarios, not shown here, the results are similar. Where the energy dependent transport from the wind bubbles to Earth can lead to an additional depletion of low energy particles depending on the observation time and distance between source and Earth.

It should be noted, that for observation times at Earth $t_\mathrm{E}> 5\,\mathrm{Myr}$ the low energy component, released if the wind ceases, could be increased. However, the latter statements neglect the turbulence and power released by supernovae exploding in the cluster. While this could provide a scenario that can still trap low-energy CRs, it is possible that the environmental conditions in the bubble cease to resemble the simple wind profile assumed in this work. In light of these complications, we limit our investigation to a relatively small age of 5 Myr during which the power content in the bubble is dominated by stellar winds \citep{Vieu-etal-2022}. 

We assume that the source switches off at the age of 5 Myrs. Therefore, the fluxes presented in fig.~\ref{fig:earth} are lower limits.

\section{Summary}
\label{sec:summary}

We have modeled CR transport and escape in time evolving wind bubble environments under the assumption of a perfect spherical symmetry. We focus on model parameters that are expected to be found in wind bubbles around young massive stellar clusters, but note that our model could be adjusted to describe ultra-fast outflows of AGN in the context of ultra-high-energy cosmic rays as discussed in \citep{Ehlert-Oikonomou-Peretti-2025}. Injecting a time continuous DSA spectrum with spectral index $\gamma=-2$ at the termination shock, we examined the escaping spectra at the forward shock position. 

We find that:
\begin{enumerate}
    \item The escaping flux can be harder than the injected one.
    \item Low energy CRs can be significantly suppressed depending on the model parameters.
\end{enumerate} 

We stress that these findings, especially the exact position of the low energy depletion, can only be predicted with the time dependent treatment discussed in this work; however, they might change, when the spherical symmetry in a more complex wind bubble model is broken. 

Hard spectra, in particular harder than those at SNRs, can help explain the CR spectrum around the knee as shown in \cite{Kaci2025} for the case of microquasars. Here, also a low-energy suppression at the level we find for Kraichnan diffusion can be sufficient to create a similar effect, provided that CRs still reach above  PeV energies. 

In the future, we will take interactions of CRs simultaneously into account and in this way predict the time dependent hadronic contributions to the neutral secondary channels in gamma rays and neutrinos. As discussed above the dynamic influence on the primary species within the first phase of the superbubble evolution is expected to be small. However, the trapped low energy particles spend more time in the region between the shocks and are expected to undergo significant interaction with an approximate energy loss time scale of  $\tau_\mathrm{pp} \approx 5 \, \left(n_\mathrm{b}/(10 \, {\rm cm^{-3}})\right)^{-1} \left(\sigma_\mathrm{pp} / (40 \, {\rm mb})\right)^{-1} \, {\rm Myr}$. Therefore, a consistent treatment of interaction is viable to predict the fate of the low energy particles.

Re-acceleration of low energy CRs at emerging shocks of SNRs and their escape into the interstellar medium, when the forward shock is ceasing are left for future work.

\begin{acknowledgments}
The authors thank the reviewer for their thorough work that helped to improve the current manuscript.

SA and LM acknowledge funding from the German Science Foundation DFG, via the Collaborative Research Center SFB1491 "Cosmic Interacting Matters - From Source to Signal". EP acknowledges economic support from INAF through “Assegni di ricerca per progetti di ricerca relativi a CTA e precursori”. Part of this work was carried out at the APC Laboratory, whose hospitality EP and LM gratefully acknowledge.
\end{acknowledgments}

\begin{contribution}
\textbf{S.~Aerdker:} Methodology, Software, Validation, Formal analysis, Writing - Original Draft, Writing - Review \& Editing, Funding acquisition; \textbf{L.~Merten:} Methodology, Software, Validation, Formal analysis, Writing - Original Draft, Writing - Review \& Editing, Visualization, Funding acquisition; \textbf{E.~Peretti:} Conceptualization, Methodology, Validation, Writing - Original Draft, Writing - Review \& Editing

\end{contribution}

\software{CRPropa \citep{CRPropa3.2, GeneralSDE},
          numpy \citep{numpy}, 
          pandas \citep{pandas}, 
          matplotlib \citep{matplotlib},
          sympy \citep{sympy}
          }

\appendix

\section{From transport equation to stochastic differential equation}

For the readers convenience the relevant steps to transform equ.~\ref{eq:transport} into a forward Fokker Planck form are shown below:
\begin{align}
    \frac{\partial f}{\partial t} &= \nabla \cdot \left(\kappa \cdot \nabla f - \mathbf{u} f\right) + \frac{p}{3}\nabla \cdot \mathbf{u} \frac{\partial f}{\partial p} + S
\end{align}
Assuming now, that $\kappa$ has only one non-vanishing element ($\kappa_{rr}=\kappa_0$, $\kappa_{ij}=0$ otherwise), that the background flow is incompressible in the relevant region between the shocks ($\nabla \cdot \mathbf{u}=0$) and given by $\mathbf{u}=u\mathbf{e}_r$, and explicitly writing the divergence in spherical coordinates, gives
\begin{align}
    \frac{\partial f}{\partial t} &=  \frac{1}{r^2} \frac{\partial}{\partial r} \left( r^2 \left(\kappa \frac{\partial f}{\partial r} - u f\right)  \right) +  S(r, p, t) \quad.
\end{align}

To avoid weighting terms \citep[see e.g.][]{GeneralSDE} we introduce a new quantity $\tilde{f} = fr^2$ and multiply the above equation by $r^2$:
\begin{align}
    \frac{\partial \tilde{f}}{\partial t} &=  \frac{\partial}{\partial r} \left( r^2 \kappa \frac{\partial f}{\partial r} - u \tilde{f}  \right) +  \tilde{S}(r, p, t) \\
    &= -\frac{\partial}{\partial r}\left(\left(u + \frac{\partial \kappa}{\partial r} + \frac{2\kappa}{r}\right)\tilde{f}\right) + \frac{1}{2}\frac{\partial^2}{\partial r^2}\left(2\kappa \tilde{f}\right) + \tilde{S}
\end{align}
This transport equation for $\tilde{f}$ does not contain any terms linear or constant in $\tilde{f}$; apart from the source term. Therefore, no weighting terms have to be taken into account, when solving the SDE. This is different, when the SDE for the distribution function $f$ itself is solved (see, e.g., the appendix of \cite{Kopp-etal-2012}).

\section{Neglecting adiabatic energy losses}
This approach is justified by the following estimation of the expected energy loss on the upstream site of the shock. The energy loss fraction is given by:
\begin{align}
    \frac{\Delta p}{p}  &= - \frac{1}{3} \int \nabla \cdot \mathbf{u}\,\mathrm{d}t\\
    &= - \frac{2v_\infty}{3} \int \frac{1}{r} \mathrm{d} t \quad .
\end{align}
Here, the we used that the flow has the constant speed $v_\infty$ in the upstream region. To further derive the energy loss we assume that the particles are concentrated around the termination shock $r(t) \approx r_\mathrm{TS}$; which is a good approximation for the majority of the particle distributions as the density decrease exponentially on the upstream site. Furthermore, the residence time in the upstream region can be approximated as $\tau_\mathrm{up} = (4\kappa_\mathrm{up})/(u_\mathrm{up} v)$, where $v$ is the particle speed and $\kappa_\mathrm{up}$ and $u_\mathrm{up}$ are the diffusion coefficient and flow speed in the upstream region.

The energy loss will be most severe for large diffusion coefficients. In the case of our Bohm diffusion scenario with large magnetization ($\epsilon_\mathrm{B}=0.1$) this gives an estimated energy loss fraction of $\Delta p / p \approx -0.01$ for a $5\,\mathrm{PeV}$ proton and on the order of $\Delta p / p \approx -10^{-8}$ at the lower energy end $5\,\mathrm{GeV}$. For simplicity we used $r^{-1}(t) = \text{max}(r^{-1}(t))|_{t\in(t_i, t_i + \tau)}$ which is the conservative assumption of the energy loss rate.

The energy loss for Kolmogorov and Kraichnan diffusion can be larger at $5\,\mathrm{PeV}$, however, these energies are not reached in the first place.

\section{Maximal energy and cut-off}
\label{Appendix: Emax}

The maximal energy is approximated based on different scale arguments as follows: 
\begin{enumerate}
    \item The maximal energy inferred by equating the upstream diffusion length, $D/u$, with the size of the upstream region itself, $R_{\rm sh}$. This leads to 
    \begin{align}
        E^\mathrm{max}_\mathrm{DSA} &= \left( \frac{3r_\mathrm{TS}u}{c}\right)^\frac{1}{2-\delta} l_0^\frac{1-\delta}{2-\delta} q B c \label{eq:Emax_DSA}
    \end{align}

    \item The energy defined by the acceleration time scale 
    \begin{align}
        E^\mathrm{max}_\mathrm{acc} &= \left(\frac{3(2-\delta)}{20}\frac{u_1^2}{c} t \right)^\frac{1}{2-\delta} l_0^\frac{1-\delta}{2-\delta} q B c \label{eq:Emax_tacc}
    \end{align}
    
\end{enumerate}

For any given time the minimal value of equs.~(\ref{eq:Emax_DSA} - \ref{eq:Emax_tacc}) is used. 

In the future, the simulations will be set up to include the acceleration process itself. This will allow to test the assumptions of the maximal energy approximation in a realistic time dependent scenario.

\section{Diffusive and advective tracer}
\label{ssec:time-scales}

\begin{figure}[htbp]
    \begin{minipage}{.49\linewidth}
        \includegraphics[width=\linewidth]{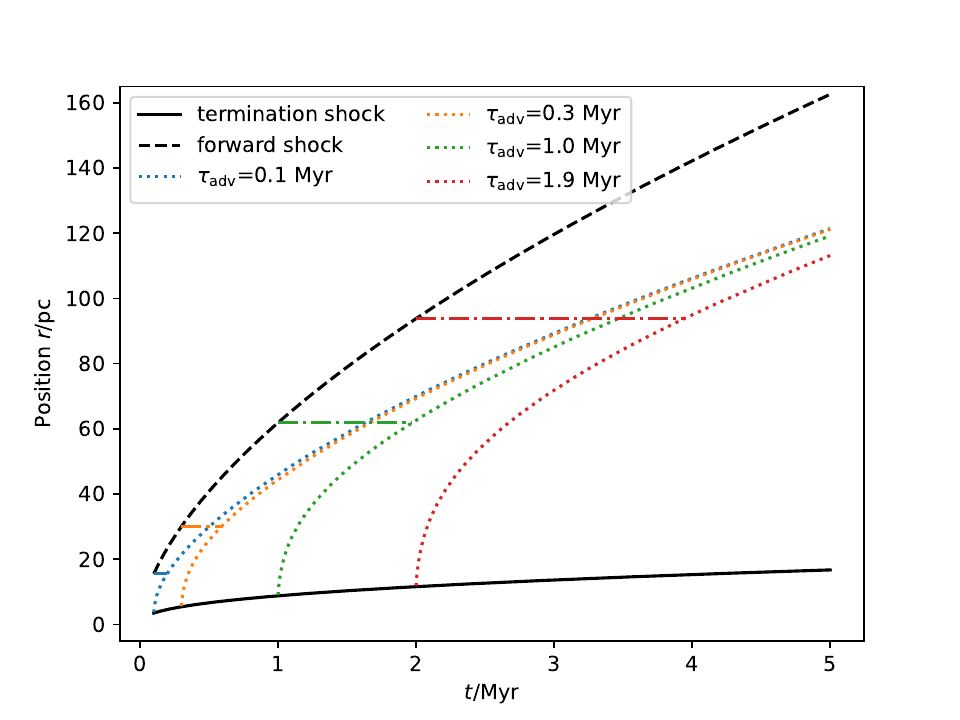}
    \end{minipage}
    \begin{minipage}{.49\linewidth}
        \includegraphics[width=\linewidth]{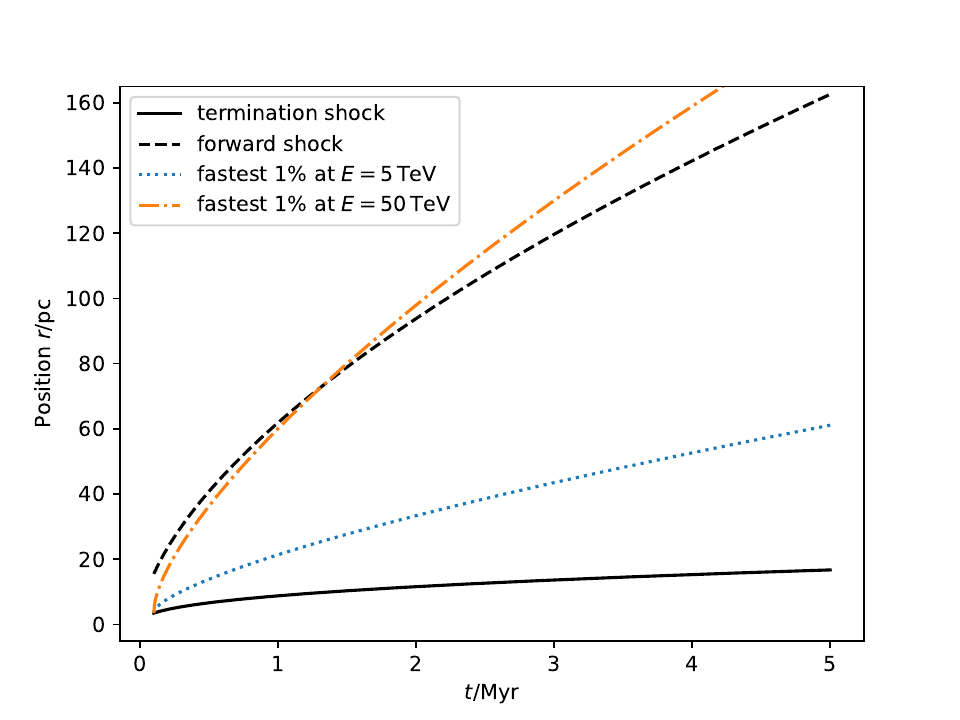}
    \end{minipage}
\caption{Time scale approximations: (Left) The positions of a purely advective particle are shown in comparison to the two shock positions (dotted lines). For comparison the time scales in the "frozen" scenario are marked with dash-dotted lines. (Right) The position of the fastest 1\% percent of purely diffusive particles at different energies.}
    \label{fig:tracer}
\end{figure}

The advection time scale derived from the "frozen" wind bubble approximation is on the order of the system age $\tau_\mathrm{adv}\approx t$. This would mean that particles injected at early times $t_\mathrm{inj}\lesssim 2.5\,\mathrm{Myr}$ would be able to leave the downstream region advectively before the end of the simulation time. However, such a behavior, which would lead to escaping particles at all energies independent of their diffusion, was not observed in our models. Therefore, we take a closer look at the advection in our time dependent model by solving equ.~\ref{eq:transport} without diffusion. In this way, we derive trajectories of advective tracer in a fully time dependent scenario. Some example trajectories are shown in fig.~\ref{fig:tracer}: It is clearly visible that, with the chosen parameters, particles cannot escape the wind bubble without diffusion. 

Similar estimates can be derived for the purely diffusive case. We calculated the the mean propagation distance of the fasted percent of all injected particles as $D_{0.01} \approx \sqrt{2 \kappa(p) t}\times 2.327$. The approximate position of these purely diffusive tracer is displayed in fig.~\ref{fig:tracer}. This approximation can be interpreted as follows: 1) Less than one percent of the particles injected at $t=0.1\,\mathrm{Myr}$ with an energy of $E<5\,\mathrm{TeV}$ will escape diffusively alone. 2) More than one percent of the particles with energies above $E<50\,\mathrm{TeV}$ will escape at a time between 1-2 Myr. 
Comparing this with the low energy depletion cut off for the same model parameters ($\alpha=0, \delta=1, \epsilon_\mathrm{B}=0.01$) it is clearly visible that the diffusion approximation alone does only provide an order of magnitude estimate of the low energy cut off.

In summary, neither the advection time estimates nor the diffusive tracer are able to correctly predict the features of the escaping flux. Only a time dependent diffusion-advection model, as in this work can do so.

\section{Spatial distribution}

\begin{figure}[htbp]
        \centering\includegraphics[width=.75\linewidth]{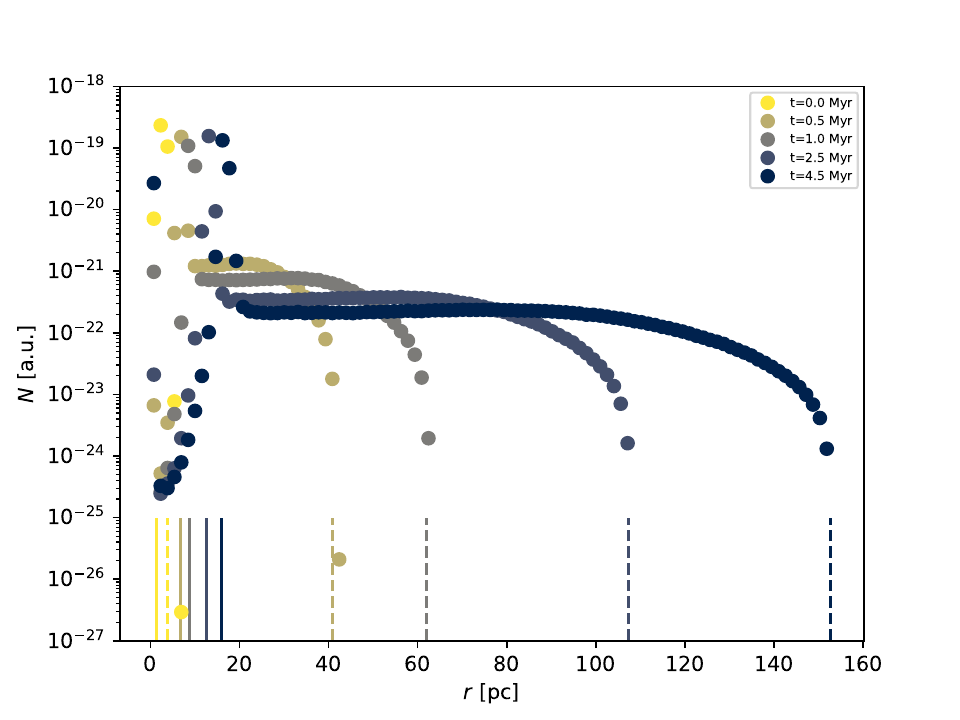}
    \caption{CR number density for a constant downstream magnetic field strength ($\alpha=0$), Kraichnan diffusion coefficient ($\delta=3/2$) and large magnetization ($\epsilon_B=0.1$). The time evolution is color coded with bright yellow corresponding to earlier and dark blue colors corresponding to later times. The positions of the termination $R_\mathrm{TS}$ and forward $R_\mathrm{FS}$ shock are marked with a solid and dashed line, respectively.}
    \label{fig:spatial}
\end{figure}

In addition to the escaping CRs, all particles within the simulation volume have been recorded several times during the simulation. This allows to reconstruct the number density of the CR distribution at selected times. Figure~\ref{fig:spatial} shows the number density $N=4\pi p^2 f$ for the same model settings as discussed above ($\alpha=-1$, $\delta=3/2$, and $\epsilon_\mathrm{B}=0.1$). The number density is highest at the current termination shock position, which is where the CRs are continuously injected. The particle density decreases sharply on the upstream side of the termination shock, which is again expected from DSA models. This validates the approximation to neglect the upstream cooling for simplicity as it would not influence the total energetics significantly. 

These features around the termination shock are observed in similar manifestations for all other parameter choices, too (see the appendix). The region between the shock looks different depending on the diffusion model and the radial evolution of the magnetic background field. For Bohm diffusion in a constant downstream magnetic field ($\delta=1$ and $\alpha=0$) lead to an almost constant density before a time dependent pile up is observed. Beyond this local maximum the density drops sharply until the forward shock is reached. The pileup is less pronounced for the cases of a decreasing downstream magnetic field strength ($\alpha<0$). 

Kolmogorov and Kraichnan diffusion do not show this pile up at all. They simply show a constant to exponentially decreasing number density, depending on the exact parameter settings. In general, Kolmogorov diffusion shows stronger radial dependence compared to Kraichnan diffusion for otherwise unchanged model parameters.

\section{Additional Plots}

\begin{figure}[htbp]
 \centering
 \includegraphics[width=\linewidth]{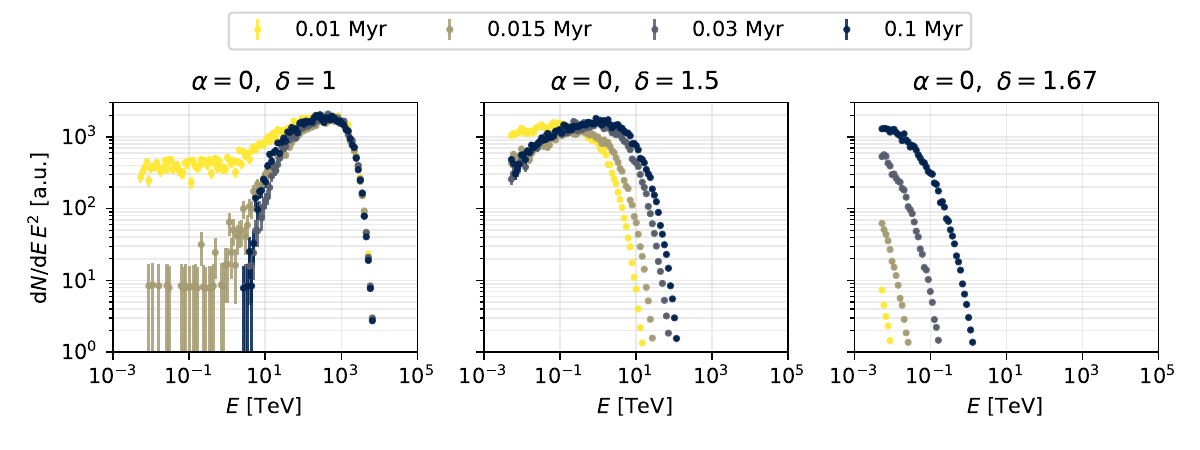}
    \caption{Escaping spectra at early times for a constant downstream magnetic field strength ($\alpha=0$) and large magnetization ($\epsilon_\mathrm{B}=0.1$) including the cut off. From left to right: Bohm, Kraichnan, and Kolmogorov diffusion.}
    \label{fig:escape_spectrum_alpha=0_early}
\end{figure}

\begin{figure}[htbp]
 \centering
 \includegraphics[width=\linewidth]{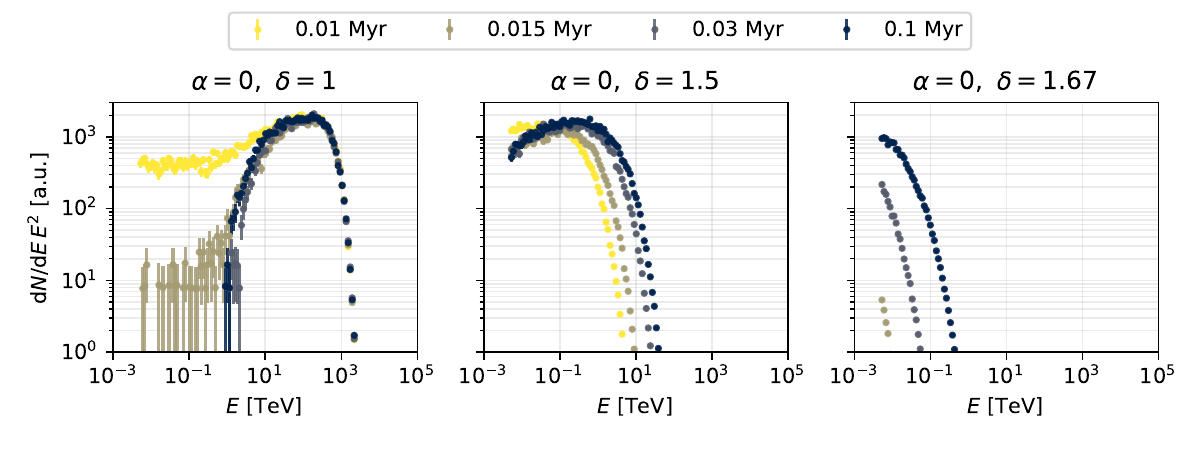}
    \caption{Escaping spectra at early times for a constant downstream magnetic field strength ($\alpha=0$) and small magnetization ($\epsilon_\mathrm{B}=0.01$) including the cut off. From left to right: Bohm, Kraichnan, and Kolmogorov diffusion.}
    \label{fig:escape_spectrum_alpha=0_early_low}
\end{figure}

Figure \ref{fig:escape_spectrum_alpha=0_early} shows the escaping spectra at earlier times than discussed in section \ref{sec:results}. It shows nicely in the case of Bohm diffusion (left plot) how the depletion at low energies is building up when the system of shocks starts to move. For the case of Kolmogorov diffusion (right) this feature the opposite effect is visible, because the maximal energy is increasing rapidly.
Figure \ref{fig:escape_spectrum_alpha=0_early_low} shows early times as well but for he case of small magnetization ($\epsilon_\mathrm{B}=0.01$).

\begin{figure}[htbp]
    \centering
    \includegraphics[width=\linewidth]{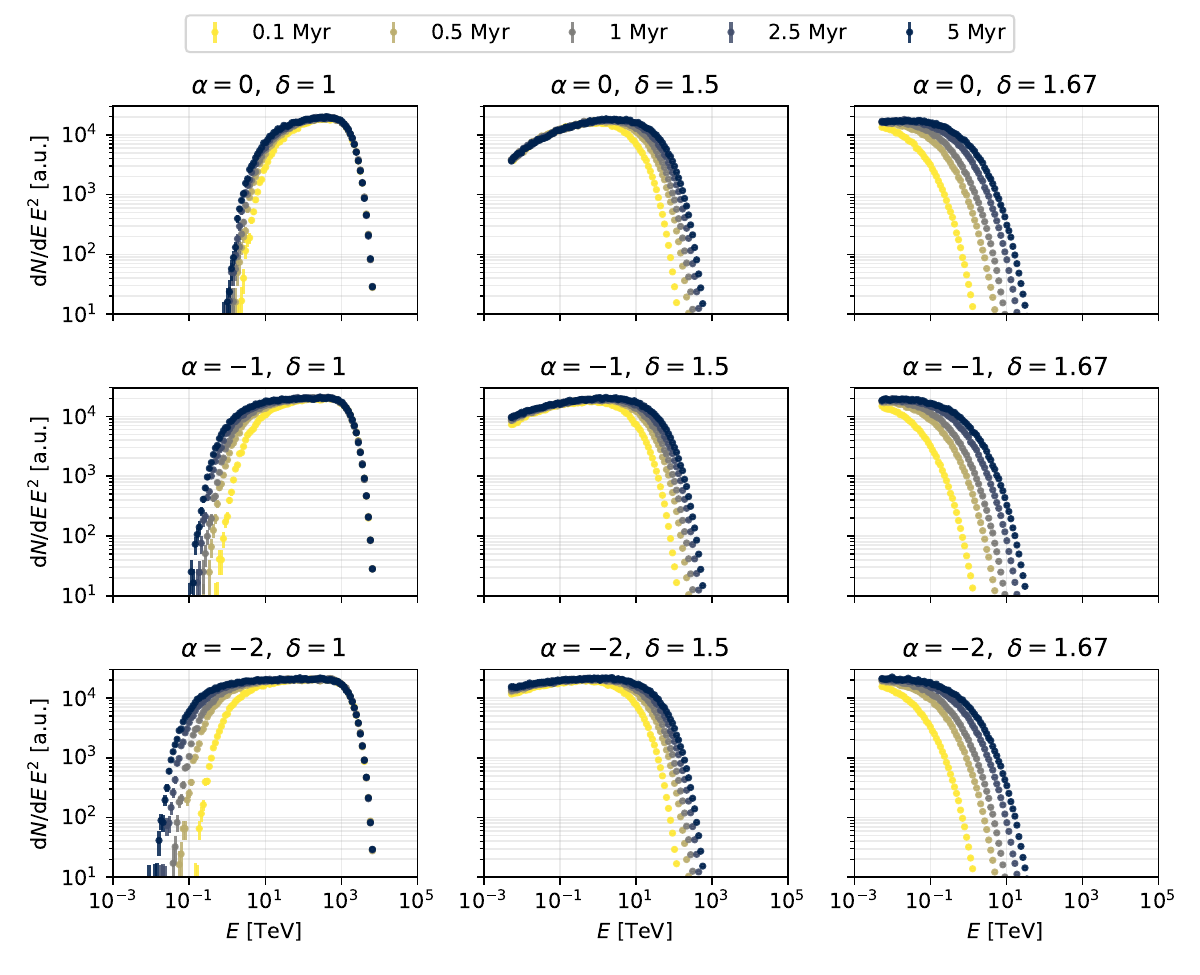}
    \caption{Escaping spectra at late times for the case of large magnetization ($\epsilon_\mathrm{B}=0.1$); cut off included. From left to right: Bohm, Kraichnan, and Kolmogorov diffusion. From top to bottom: no ($\alpha=0$), medium ($\alpha=-1$), and strong ($\alpha=-2$) radial dependence of the magnetic field.}
    \label{fig:late-time-overview}
\end{figure}

\begin{figure}[htbp]
    \centering
    \includegraphics[width=\linewidth]{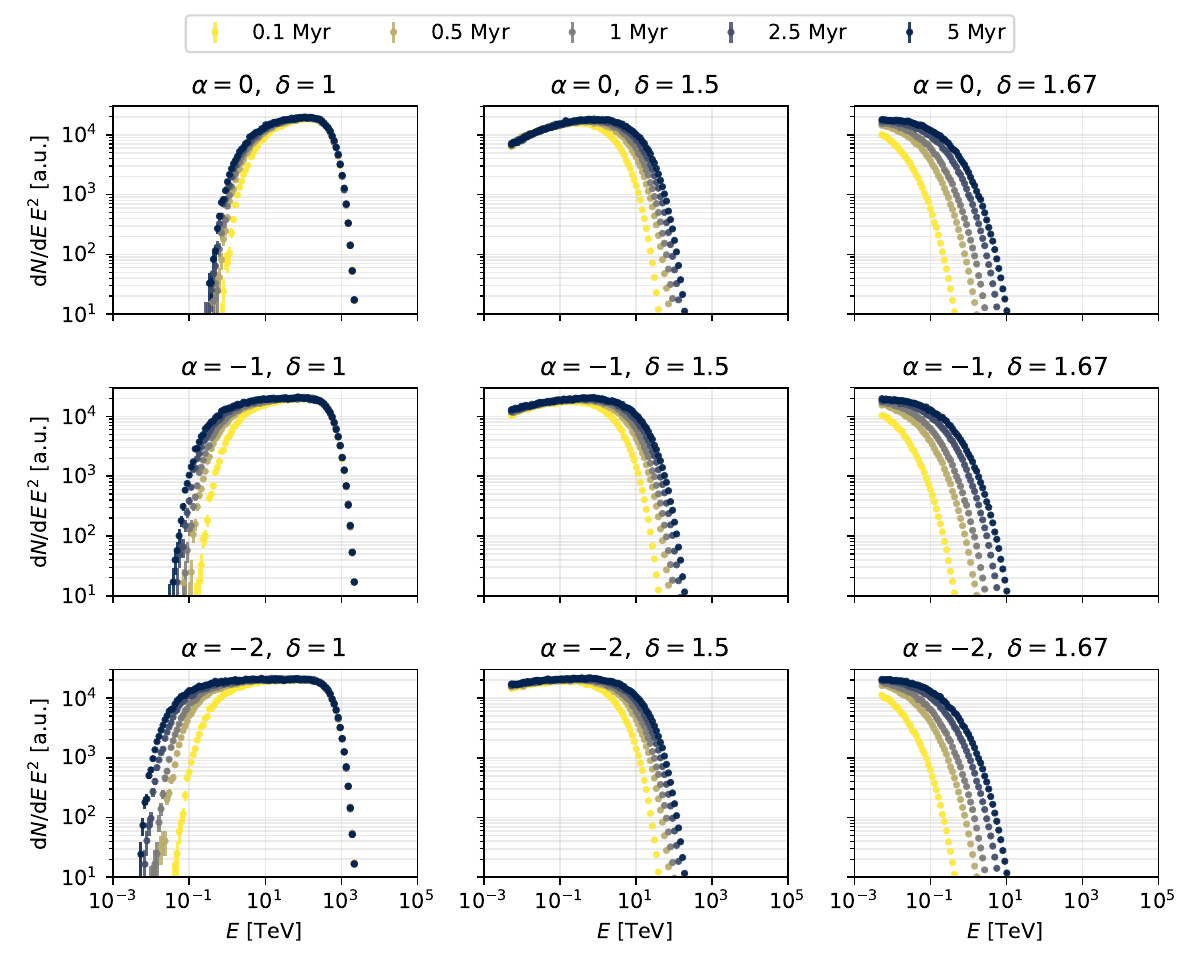}
    \caption{Escaping spectra at late times for the case of small magnetization ($\epsilon_\mathrm{B}=0.01$); cut off included. From left to right: Bohm, Kraichnan, and Kolmogorov diffusion. From top to bottom: no ($\alpha=0$), medium ($\alpha=-1$), and strong ($\alpha=-2$) radial dependence of the magnetic field.}
    \label{fig:late-time-overview_low}
\end{figure}

Figure \ref{fig:late-time-overview} shows a compilation of all escaping spectra at late times for the case of strong magnetization $\epsilon_\mathrm{B}=0.1$; see fig.~\ref{fig:late-time-overview_low} for the small magnetization. It clearly shows that a stronger radial dependence of the magnetic field --- larger diffusion coefficients --- lead to a suppression of the depletion for Bohm and Kraichnan diffusion (left and center column). In the case of Bohm diffusion the mean free path of the particles is so large independent of the radial magnetic field dependence that almost no difference in the spectra is visible (right column).

For a fixed radial dependence (line wise comparison), the suppression is largest for Bohm and smallest for Kolmogorov diffusion. However, for the time dependence of the maximal energy the opposite is true.

Additional figures and the raw data of the plots are available on reasonable request to the authors.

\newpage
\bibliography{sample701}{}
\bibliographystyle{aasjournalv7}

\end{document}